\documentclass[12pt]{article}
\usepackage{epsfig,amsmath}

\title{\bf Rotating Relativistic Thin Disks}

\author{{\it Guillermo A. Gonz\'{a}lez}\thanks{e-mail: guillego@uis.edu.co}\\
Escuela de F\'{\i}sica	\\
Universidad Industrial de Santander	\\
A.A. 678, Bucaramanga, Colombia	\\
{\small and }\\
{\it Patricio S. Letelier}\thanks{e-mail: letelier@ime.unicamp.br}	\\
Departamento de Matem\'{a}tica Aplicada	\\
Universidade Estadual de Campinas	\\
13081-970, Campinas, S.P., Brazil}

\date{ }

\begin{document}

\maketitle

\begin{abstract}

Two families of  models of rotating  relativistic disks based
on Taub-NUT and Kerr metrics are constructed using the  well-known 
``displace, cut and reflect" method.  We find  that for  disks built
from a  generic stationary axially symmetric metric  the ``sound 
velocity", $(pressure/density)^{1/2}$, is equal to the geometric
mean of the prograde and retrograde  geodesic circular velocities of test
 particles  moving on the
disk. We also found that  for generic disks we can have zones with
heat flow. For the two families of models studied the boundaries 
that
separate the zones with and without  heat flow are  not stable against
radial perturbations (ring formation).
\vspace{0.4cm}

\end{abstract}

\newpage

\section{Introduction}

Axially symmetric solutions of Einstein field equations 
corresponding to
disk like configurations of matter are of great astrophysical
 interest
and have been extensively studied. These solutions can be 
static or
stationary and with or without radial pressure.  Solutions 
for static
disks without radial pressure were first studied by Bonnor 
and Sackfield
\cite{BOSA}, and Morgan and Morgan \cite{MM1}, and with 
radial pressure
by Morgan and Morgan \cite{MM2}, and, in connection with
 gravitational
collapse, by Chamorro, Gregory and Stewart \cite{CHGS}. Disks with radial
 tension has been recently considered in \cite{gole}. Also models 
  of disks with, electric fields \cite{lzb},
 magnetic fields \cite{hot}, and both  magnetic and electric fields have
  been introduced recently \cite{hote}. Using a
  solitonic technique Neugebauer and Meinel found solutions
  representig rigidly rotating disks of dust \cite{nm}. Several classes of 
exact solutions of the Einstein field equations
that represent disks  were obtained by  different authors \cite{LO}-\cite{BLP}.

 The  stability of static  models  with no radial pressure can be
 explained by either assuming the
existence of hoop stresses or that the particles on the disk plane
move under the action of their own gravitational field in such a
way that as many particles move clockwise as counterclockwise. This
last interpretation is frequently made since it can be invoked to
mimic true rotational effects. Even though, this interpretation can be 
seen as a device, there are observational evidence of disks 
made of streams of  rotating and counter-rotating matter \cite{counter}.

For disks without tension, i.e., usual fluid disks, the rotation 
is a necessary ingredient to have stability.  Despite its relevance
the literature on exact relavistic disks solutions to the Einstein field
equations is scarce, we only found 
a discussion of a rotating disk as a source of Kerr metric  \cite{BL}.

The aim of  this work is to study  models of rotating 
relativistic disks in some 
detail. The models are based in Taub-NUT and Kerr metrics, that are some
of the simplest axially symmetric stationary solutions of the vacuum
Einstein equations \cite{KSM}. To construct the disk
solutions we use the well-known  ``displace, cut and reflect" method. 
 
 We find  that for  disks built from a generic stationary axially symmetric
 metric  the ``sound velocity", $ (pressure/density)^{1/2}  $, is equal
 to the geometric mean of the prograde and retrograde 
 geodesic circular velocities of a test particle moving 
 on the disk.   We also found that the
generic disk  may have zones with heat flow. For the two families of
  models studied the boundaries that separates the zones with and without 
  heat flow are  not stable against radial perturbations (ring formation).

 In  Sec. II we study the generation of a disk model using 
the ``displace, cut and reflect" 
method from a vacuum solution
of the Einstein field equations for a stationary axially symmetric metric, in
particular, we study:  the general form of the associated  energy-momentum
tensor,   disk velocity, and angular momentum. In the next section, Sec.
III, we discuss the relation between the geodesic circular velocities 
 on the disk and the pressure and density.  In Sec. IV a class of models
 of rotating disks based on the Taub-NUT metric is presented. In particular,
 we study the  apparition of heat flow and stability against radial
 perturbations. In Sec. V, another class of models of
 rotating disks is presented, this time based in the Kerr metric. Also
 the  stability against radial perturbation is considered. The material
 presented here is complementary to the work of reference \cite{BL}. In the last section,
 Sec VI , we summarize our main results and made some comments.

\section{Rotating Relativistic Disks}

In this section we present a summary of the main quantities associated to
the disk, we follow closely our reference \cite{gole}.
A sufficiently general metric for our purposes is Weyl-Lewis-Papapetrou line 
element, 
\begin{equation}
ds^2 = - \ e^{2 \Phi} (dt + {\cal A} d\varphi)^2 \ + \ e^{- 2 \Phi} [r^2
d\varphi^2 + e^{2 \Lambda} (dr^2 + dz^2)], \label{eq:met}
\end{equation}
where $\Phi$, $\Lambda$ and ${\cal A}$ are functions of $r$ and $z$ only.
 Assuming  the existence of the second derivatives of the functions 
 $\Phi$, $\Lambda$ and ${\cal A}$,
 the Einstein vacuum equations for this metric are equivalent to,
\begin{subequations}
\begin{eqnarray}
\Phi_{,rr} + \frac{\Phi_{,r}}{r} + \Phi_{,zz} + \frac{e^{4 \Phi}}{2 r^2} (
{\cal A}_{,r}^2 + {\cal A}_{,z}^2 ) & = & 0 , \label{eq:ei1} \\
\	\nonumber	\\
{\cal A}_{,rr} - \frac{{\cal A}_{,r}}{r} + {\cal A}_{,zz} + 4 (\Phi_{,r} {\cal
A}_{,r} + \Phi_{,z} {\cal A}_{,z} ) & = & 0 , \label{eq:ei2}
\end{eqnarray}\begin{eqnarray}
\Lambda_{,r} & = & r ( \Phi_{,r}^2 - \Phi_{,z}^2 ) - \frac{e^{4 \Phi}}{4 r} (
{\cal A}_{,r}^2 - {\cal A}_{,z}^2 ) , \label{eq:ei3} \\
\	\nonumber	\\
\Lambda_{,z} & = & 2 r \Phi_{,r} \Phi_{,z} - \frac{e^{4 \Phi}}{2 r} {\cal
A}_{,r} {\cal A}_{,z} . \label{eq:ei4}
\end{eqnarray}
\end{subequations}

Now if the first derivatives of the metric tensor are not continuous
on the  plane $z=0$ with discontinuity functions,
$$
b_{ab} \ = \ g_{ab,z}|_{_{z = 0^+}} \ - \ g_{ab,z}|_{_{z = 0^-}},
$$
the Einstein equations yield 
an  energy-momentum tensor (EMT) $T_{ab} = Q_{ab} \ \delta(z)$,
where $\delta(z)$ is the usual Dirac function with support
 on the disk and
$$
Q^a_b = \frac{1}{2}\{b^{az}\delta^z_b - b^{zz}\delta^a_b + g^{az}b^z_b -
g^{zz}b^a_b + b^c_c (g^{zz}\delta^a_b - g^{az}\delta^z_b)\}.
$$
is the distributional energy-momentum tensor. The ``true'' surface
energy-momentum tensor of the disk can be written as $S_{ab} = e^{\Lambda -
\Phi} \ Q_{ab}$.

For the metric (\ref{eq:met}) we obtain
\begin{subequations}
\begin{eqnarray}
&S_{00} &= \ 2 e^{3\Phi - \Lambda} \left[ 2 \Phi_{,z} - \Lambda_{,z}
\right], \label{eq:emt1}	\\
 &	&	\nonumber		\\
&S_{01} &= \ e^{3\Phi - \Lambda} \left[ 4 {\cal A} \Phi_{,z} - 2 {\cal A}
\Lambda_{,z} + {\cal A}_{,z}  \right] \label{eq:emt2}	\\
&	&	\nonumber		\\
&S_{11} &= \ 2 e^{3\Phi - \Lambda} \left[ ( r^2 e^{- 4 \Phi} - {\cal A}^2 )
\Lambda_{,z} + 2 {\cal A}^2 \Phi_{,z} + {\cal A}{\cal A}_{,z} \right]
\label{eq:emt3}
\end{eqnarray}
\end{subequations}
where all the quantities are evaluated at $z = 0^+$.

The eigenvalue problem for the energy-momentum tensor (\ref{eq:emt1}) -
(\ref{eq:emt3}),
\begin{equation}
S^a_b \ \xi^b \ = \lambda \ \xi^a,
\end{equation}
has the solutions
\begin{equation}
\lambda_\pm \ = \frac{1}{2} \left( \ T \pm \sqrt{D} \ \right),
\end{equation}
and $\lambda_r = \lambda_z = 0$,
where
\begin{equation}
T = S^0_0 \ + \ S^1_1 ,	\qquad D = ( S^1_1 - S^0_0 )^2 + 4 \ S^0_1 \ S^1_0 .
\end{equation}

We can write 
\begin{equation}
g_{ab} \ = \ - V_a V_b + W_a W_b + X_a X_b + Y_a Y_b , \label{eq:metdia}
\end{equation}
and the canonical form of the EMT,
\begin{equation}
S_{ab} \ = \ \sigma V_a V_b + P W_a W_b + K \left( V_a W_b + W_a V_b \right) ,
\label{eq:emtdia}
\end{equation}
with a orthonormal basis
\begin{subequations}
\begin{eqnarray}
V^a & = & N_0 ( 1, \Omega, 0, 0 ),		\\
	&	&	\nonumber		\\
W^a & = & N_1 ( \Delta, 1, 0, 0 ), 		\\
	&	&	\nonumber		\\
X^a & = & e^{\Phi - \Lambda} ( 0, 0, 1, 0 ),	\\
	&	&	\nonumber		\\
Y^a & = & e^{\Phi - \Lambda} ( 0, 0, 0, 1 ),
\end{eqnarray}
\end{subequations}
where $N_0$ and $N_1$ are normalization factors, and
\begin{equation}
\Omega \ = \ \left\{ \begin{array}{ccc}
(\lambda_- - S^0_0)/S^0_1& , & D \geq 0 \ , \\
	&	&	\\
(S^1_1- S^0_0)/2 S^0_1 & , & D \leq 0 \ ,\\
\end{array} \right.
\end{equation}
\begin{equation}
\Delta \ = \ \left\{ \begin{array}{ccc}
(\lambda_+ - S^1_1)/S^1_0 & , & D \geq 0 \ , \\
	&	&	\\
0 & , & D \leq 0 \ ,\\
\end{array} \right.
\end{equation}

The energy density, the azimuthal pressure, and  the heat flow function are,
respectively,
\begin{equation}
\sigma \ = \ \left\{ \begin{array}{ccc}
- \lambda_- & , & D \geq 0 \ , \\
	&	&	\\
- T/2 & , & D \leq 0 \ ,\\
\end{array} \right.
\end{equation}
\begin{equation}
P \ = \ \left\{ \begin{array}{ccc}
\lambda_+ & , & D \geq 0 \ , \\
	&	&	\\
T/2 & , & D \leq 0 \ ,\\
\end{array} \right.
\end{equation}
\begin{equation}
K \ = \ \left\{ \begin{array}{ccc}
0 & , & D \geq 0 \ , \\
	&	&	\\
\sqrt{- D}/2 & , & D \leq 0 \ .\\
\end{array} \right.
\end{equation}


The orthonormal basis $\{ V^a , W^a , X^a , Y^a \}$ is 
comoving with the disk.
Thus the time-like vector $V^a$ will define the velocity vector of
the disk:
\begin{equation}
V^a \ = \ \left( V^0 , V^1 , 0 , 0 \right) \ =  \ V^0 \left( 1 , \Omega , 0 , 0
\right) ,
\end{equation}
where
\begin{equation}
V^0 \ = \ \frac{e^{- \Phi}}{\sqrt{1 - V^2}},
\end{equation}
and
\begin{equation}
V \ = \ \sqrt{\frac{g_{11} \Omega^2 + 2 g_{01} \Omega}{- g_{00}}}
\end{equation}
is the tangential velocity of the disk. 

The specific angular momentum of a particle of the disk, with mass $\mu$, is
given by
\begin{equation}
h \ = \ \frac{p_\varphi}{\mu} \ = \ g_{\varphi a} V^a .
\end{equation}
Thus,
\begin{equation}
h \ = \ \frac{g_{11} \Omega + g_{01}}{\sqrt{- g_{00} - 2 g_{01} \Omega - g_{11}
\Omega^2 }} .
\end{equation}
A condition of stability  under radial perturbations is
\begin{equation}
\frac{d(h^2)}{dr} = 2 h \frac{dh}{dr} \ > \ 0 \ ,
\end{equation}
For $h > 0$, we have stability when the specific angular momentum is an
increasing function of $r$.
This criteria is an extension of  Rayleigh
criteria of stability of a fluid at rest in a gravitational field, see
 for instance \cite{LLfluids}.
 
\section{The counter-rotating Model}

In this section we analyze the model of disks made of particles moving in opposite
directions. When the discriminant $D$ is positive  there is not heat flow. Furthermore,
when $P > 0$  an observer comoving with the disks can consider the energy-momentum 
tensor as representing
two streams of collisionless particles that circulate in opposite directions.

Let be $u^a = ( u^0 , u^1 , 0 , 0 ) = u^0 ( 1 , \omega , 0 , 0 )$ the velocity
vector of the stream. The  angular velocity $\omega$ can be obtained from
the geodesic equation for a test particle, we get
\begin{equation}
g_{11,r} \omega^2 + 2 g_{01,r} \omega + g_{00,r} = 0 , \label{eq:ecge}
\end{equation}
with solutions,
\begin{equation}
\omega_\pm \ = \ \frac{- g_{01,r} \pm \sqrt{g_{01,r}^2 - g_{00,r}
g_{11,r}}}{g_{11,r}} .
\end{equation}
Therefore, in general, the two streams circulate with different
 velocities.

We can compute the tangential velocity of the streams by projecting the
velocity vector $u^a$ onto the comoving tetrad, ${e_{\hat a}}^b$ = $\{ V^b ,
W^b , X^b , Y^b \}$,
\begin{equation}
u^{\hat a} \ = \ {e^{\hat a}}_b u^b = \eta^{{\hat a}{\hat c}} e_{{\hat c}b}
u^b  .
\end{equation}
 We get,
\begin{equation}
U_\pm \ = \ \left| \frac{u^{\hat 1}}{u^{\hat 0}} \right| = \left| \frac{W_0 +
W_1 \omega_\pm}{V_0 + V_1 \omega_\pm} \right| \ .
\end{equation}
By  performing  the product of $U_+$ and $U_-$ we obtain
\begin{equation}
U_+ U_- \ = \ \left| \frac{{W_0}^2 + W_0 W_1 (\omega_+ + \omega_-) + {W_1}^2
\omega_+ \omega_- }{{V_0}^2 + V_0 V_1 (\omega_+ + \omega_-) + {V_1}^2 \omega_+
\omega_- } \right| .
\end{equation}
From (\ref{eq:ecge}) we have the relations,
\begin{equation}
\omega_+ + \omega_- \ = \ - \frac{2 g_{01,r}}{g_{11,r}} , \qquad \omega_+
\omega_- \ = \ \frac{g_{00,r}}{g_{11,r}} .
\end{equation}
Thus,
\begin{equation}
U_+ U_- \ = \ \left| \frac{g_{11,r} {W_0}^2 - 2 g_{01,r} W_0 W_1 + g_{00,r}
{W_1}^2 }{g_{11,r} {V_0}^2 - 2 g_{01,r} V_0 V_1 + g_{00,r} {V_1}^2 } \right| .
\end{equation}

By using (\ref{eq:metdia}) and (\ref{eq:emtdia}), we can write
\begin{equation}
U_+ U_- \ = \ \left| \frac{ A + \sigma B }{ A - P B } \right| ,
\end{equation}
where
\begin{subequations}
\begin{eqnarray}
A & = & g_{11,r} S_{00} - 2 g_{01,r} S_{01} + g_{00,r} S_{11} , \\
	&	&	\nonumber	\\
B & = & g_{00} g_{11,r} - 2 g_{01} g_{01,r} + g_{00,r} g_{11} .
\end{eqnarray}
\end{subequations}
Using the Einstein equations (\ref{eq:ei1}) - (\ref{eq:ei4}) and the
expressions (\ref{eq:emt1}) - (\ref{eq:emt3}) for the energy-momentum tensor 
(with $K = 0$)  we can show that
\begin{equation}
A \ = \ - 2 r ( S_0^0 + S_1^1).
\end{equation}
Also,  using the metric (\ref{eq:met}) we get,
\begin{equation}
B \ = \ - 2 r .
\end{equation}
Hence,  assuming  that $P > 0$ and $\sigma > 0$, we obtain

\begin{eqnarray}
U_+ U_- &=&\left| \frac{S_0^0 + S_1^1 + \sigma}{S_0^0 + S_1^1 - P} \right| \\
&=&\frac{P}{\sigma} . \label{eq:sigpre}
\end{eqnarray}

In other words,  for  disks built from a generic stationary axially symmetric
 metrics  the ``sound velocity" $(P/\sigma)^{1/2} $ is equal
 to the geometric mean of the prograde and retrograde 
 geodesic circular velocities of a test particle moving on the disk.  

\section{Taub-NUT Disks}

The simplest stationary axially symmetric solution of the Einstein equations is 
the Taub-NUT solution that can be written as (\ref{eq:met}) with
\begin{subequations}
\begin{eqnarray}
\Phi & = & \frac{1}{2} \ln \left[ \frac{ x^2  - 1 }{ x^2 + 2 p x + 1 }
\right] , \\
	&	&	\nonumber	\\
\Lambda & = & \frac{1}{2} \ln \left[ \frac{ x^2 - 1 }{ x^2 - y^2 }
\right] , \\
	&	&	\nonumber	\\
{\cal A} & = &  2 \alpha q  y  , 
\end{eqnarray}
\end{subequations}
where $p = m/\alpha$ , $q =  l/\alpha$ , with $\alpha^2 = m^2 + l^2$, so that
$p^2 + q^2 = 1$. $m$ and $l$ are the mass and the NUT parameter, respectively.
$x$ and $y$ are the prolate spheroidal coordinates, given by
\begin{subequations}
\begin{eqnarray}
2 \alpha x & = & \sqrt{ r^2 + ( z + b + \alpha )^2 } + \sqrt{ r^2 + ( z + b -
\alpha )^2 } , \label{eq:corp1}	\\
	&	&	\nonumber	\\
2 \alpha y & = & \sqrt{ r^2 + ( z + b + \alpha )^2 } - \sqrt{ r^2 + ( z + b
- \alpha )^2 } , \label{eq:corp2}
\end{eqnarray}
\end{subequations}
with $b > \alpha > 0$. Note that we have  displaced the origin of the $z$-axis
in $b$.
From the above expressions we can compute the trace ($T$) and discriminant 
($D$)
of the energy-momentum tensor. We found  ${\tilde T} = \alpha T$
and ${\tilde D} = \alpha^2 D$, with
\begin{subequations}
\begin{eqnarray}
{\tilde T} & = & - \frac{ 4 {\bar y} [ {\bar y}^2 ( 2 {\bar x}^3 + 3 p {\bar
x}^2 - p ) + {\bar x} ( p {\bar x}^3 - 3 p {\bar x} - 2 ) ] }{ [ ({\bar x}^2 -
{\bar y}^2 ) ( {\bar x}^2 + 2 p {\bar x} + 1 ) ]^{3/2} }	\\
	&	&	\nonumber	\\
{\tilde D} & = & \frac{ 16 [ {\bar y}^2 ( p {\bar x}^2 + 2 {\bar x} + p )^2 -
q^2 {\bar x}^2 ( {\bar x}^2 - 1 ) ( 1 - {\bar y}^2 ) ] }{ ( {\bar x}^2 - {\bar
y}^2 ) ( {\bar x}^2 + 2 p {\bar x} + 1 )^3 } .
\end{eqnarray}
\end{subequations}
${\bar x}$ and ${\bar y}$ are given by
\begin{subequations}
\begin{eqnarray}
2 {\bar x} & = & \sqrt{ {\tilde r}^2 + ( \kappa + 1 )^2 } + \sqrt{ {\tilde
r}^2 + ( \kappa - 1 )^2 } , \\
	&	&	\nonumber	\\
2 {\bar y} & = & \sqrt{ {\tilde r}^2 + ( \kappa + 1 )^2 } - \sqrt{ {\tilde
r}^2 + ( \kappa - 1 )^2 } ,
\end{eqnarray}
\end{subequations}
where ${\tilde r} = r/\alpha$ and $\kappa = b/\alpha$.

Even though  we have an exact solution and all the relevant quantities of the
disk can be explicitly computed the resulting expressions are cumbersome
and not very illuminating,  the adoption of the graphic method is more 
adequate in this case. The first quantity to be considered is the discriminant D
that will give us the canonical form of the EMT.

In Fig. \ref{fig:disnut} we show $\tilde D$ for  Taub-NUT disks with different
values of $p$ and $\kappa$. First we plot $\tilde D$ for  disks with $\kappa =
1.7$ and $p = 0.9$, $0.7$, $0.5$, $0.3$, and $0.1$. Then we plot $\tilde D$ for
 disks with $p = 0.8$ and $\kappa = 2$, $2.6$, $3.2$, $3.8$, and $4.4$. We also
computed $\tilde D$ for many other values of $p$ and $\kappa$. We found
that  in all the cases when $p \neq 1$, $D$ is not positive definite and
it has only one root ${\tilde r}_0 > 0$. Therefore the Taub-NUT 
disks always have  heat flow beginning at ${\tilde r} = {\tilde r_0}$.

In Fig \ref{fig:depenut} we depict the density ${\tilde \sigma} = \alpha \sigma$
and the pressure ${\tilde P} = \alpha P$ for  Taub-NUT disks with different
values of $p$ and $\kappa$. First we plot $\tilde \sigma$ and $\tilde P$
(scaled by a factor $10$) for  Taub-NUT disks with $\kappa = 1.4$ and $p = 1$,
$0.8$, and $0.6$ as functions of $\tilde r = r/m$. Then we plot $\tilde \sigma$
and $\tilde P$ (scaled by a factor $10$) for  Taub-NUT disks with $\kappa = 1.7$
and $p = 1$, $0.8$, and $0.6$ as functions of $\tilde r = r/m$. We also computed
$\tilde \sigma$ and $\tilde P$ for many other values of $p$ and $\kappa$,
in all the cases, we obtain a similar behaviour. The density $\tilde \sigma$ is
always positive, ${\tilde \sigma} > 0$; whereas the pressure becomes negative
(tension)  for a value of ${\tilde r} < {\tilde r}_0$. We can
also see that $\tilde \sigma$ and $\tilde P$ presents a non smooth behaviour at
${\tilde r} = {\tilde r}_0$. For ${\tilde r} > {\tilde r}_0$  we have
 heat flow.

The heat flow function $\tilde K = \alpha K$ is represented 
in Fig. \ref{fig:calnut}.
We plot $\tilde K$ for $\kappa = 1.4$ (upper two curves) and $\kappa = 1.7$ (lower
two curves), with $p = 0.8$ and $0.6$ as functions of $\tilde r$. Also in Fig.
\ref{fig:calnut}, in order to see the change of behaviour of $\tilde \sigma$
and $\tilde P$ at ${\tilde r} = {\tilde r}_0$ and the relation between $\tilde
\sigma$ and ${\tilde K}$, we plot $\tilde \sigma$, $\tilde P$ and $\tilde K$
for $\kappa = 1.1$ and $p = 0.8$ in the interval $2.5 \leq {\tilde r} \leq 3$.
We can see that ${\tilde K} > {\tilde \sigma}$ for ${\tilde r} \geq {\tilde
r}_1 > {\tilde r}_0$. Thus, there is not causal propagation of heat for
${\tilde r} > {\tilde r}_1$.

In Fig. {\ref{fig:vemanut} we show the  tangential disk velocity $V$ and
the  disk angular momentum $h$ for Taub-NUT disks with $p = 0.8$ and
$\kappa = 1.1$, $1.4$, $1.7$ and $2$, as functions of $\tilde r$. 
We see that there is a strong change in
the slope  of $V$ and $h$ at ${\tilde r} = {\tilde r}_0$. That means that
there is a strong instability  at this value of ${\tilde r}$. We
also show the stream angular momenta $h_+$ and $h_-$ in order to compare with
the counter-rotating model. We also found  instability, but in a different place.
We computed
also $V$ and $h$ for a wide range of the parameters $p$ and $\kappa$, we
 found always
the same   behaviour.

In order to compare the counter-rotating model of the disk with the true disk
rotation, we plot in Fig. \ref{fig:vcr} the tangential velocities of the
counter-rotating streams, $U_+$ (top) and $U_-$, and the product $U_+ U_-$ (bottom
full line) and
 the $P/\sigma$ (full  dots) for $\kappa = 1.1$ and $p = 0.8$, as
functions of $\tilde r$. We can see the exact matching of the two curves, $U_+
U_-$ and $P/\sigma$, in perfect agreement with (\ref{eq:sigpre}). Also in Fig.
\ref{fig:vcr}, in order to show the behavior of the counter-rotating model for
the different disks models, we plot the $P/\sigma$ relation for $\kappa = 1.4$
(upper curves) and $\kappa = 1.7$ (lower curves), with $p = 1$, $0.8$, and $0.6$
as functions of $\tilde r$.

\section{Kerr Disks}

In this section we use the ``displace, cut, and reflect" method to built disk 
solutions using the Kerr solution, that  can be written as (\ref{eq:met}) with
\begin{subequations}
\begin{eqnarray}
\Phi & = & \frac{1}{2}\ln \left[ \frac{p^2 x^2 + q^2 y^2 - 1}{ ( p x + 1 )^2 +
q^2 y^2 } \right] , \\
	&	&	\nonumber	\\
\Lambda & = & \frac{1}{2} \ln \left[ \frac{p^2 x^2 + q^2 y^2 - 1}{p^2
(x^2 - y^2)} \right] ,	\\
	&	&	\nonumber	\\
{\cal A} & = & \frac{ 2 \alpha q}{p} \left[ \frac{ (1 - y^2)( p x + 1)}{
p^2 x^2 + q^2 y^2 - 1} \right]  ,
\end{eqnarray}
\end{subequations}
where $p = \alpha/m$ , $q =  a/m$ , with $\alpha^2 = m^2 - a^2$, so that $p^2 +
q^2 = 1$. $m$ and $a$ are the mass and the Kerr parameter, respectively. $x$
e $y$ are, again, the prolate spheroidal coordinates, given by
(\ref{eq:corp1}) and (\ref{eq:corp2}).

With the above expressions we can compute the trace and the discriminant
of the energy-momentum tensor, that now  can be written as ${\tilde T} = m T$ and
${\tilde D} = m^2 D$, with
\begin{subequations}
\begin{eqnarray}
{\tilde T}  & = & \frac{4 {\bar x} {\bar y} ( {\bar x}^2 - 1 ) (1 - {\bar y}^2
) }{ p ( {\bar x}^2 - {\bar y}^2 )^2 ( p^2 {\bar x}^2 + q^2 {\bar y}^2 - 1 )}
	\\
	&	&	\nonumber	\\
& - & 2 {\bar y} \left\{ \frac{ 2 q^2 {\bar x} ( p {\bar x} + 1 ) ( 1 - {\bar
y}^2 ) + p ( {\bar x}^2 - 1 ) [ ( p {\bar x} + 1 )^2 - q^2 {\bar y}^2 ] }{ p (
{\bar x}^2 - {\bar y}^2 ) ( p^2 {\bar x}^2 + q^2 {\bar y}^2 - 1 ) [ ( p {\bar
x} + 1 )^2 + q^2 {\bar y}^2 ] } \right\} \nonumber
\end{eqnarray}
\begin{eqnarray}
{\tilde D}  & = & 4 {\bar y}^2 \left\{ \frac{ 2 q^2 {\bar x} ( p {\bar x} + 1 )
( 1 - {\bar y}^2 ) + p ( {\bar x}^2 - 1 ) [ ( p {\bar x} + 1 )^2 - q^2 {\bar
y}^2 ] }{ p ( {\bar x}^2 - {\bar y}^2 ) ( p^2 {\bar x}^2 + q^2 {\bar y}^2 - 1 )
[ (p {\bar x} + 1 )^2 + q^2 {\bar y}^2 ] } \right\}^2 \nonumber \\
	&	&	\nonumber	\\
& - & \frac{ 16 q^2 {\bar y}^2 ( {\bar x}^2 - 1 ) ( 1 - {\bar y}^2 ) [ 3 p^2
{\bar x}^2 + 4 p {\bar x} - q^2 {\bar y}^2 + 1 ]^2 }{ p^2 ( {\bar x}^2 - {\bar
y}^2 )^2 ( p^2 {\bar x}^2 + q^2 {\bar y}^2 - 1 )^3 [ ( p {\bar x} + 1 )^2 + q^2
{\bar y}^2 ] } ,
\end{eqnarray}
\end{subequations}
where
${\bar x}$ and ${\bar y}$ are given by
\begin{subequations}
\begin{eqnarray}
2 p {\bar x} & = & \sqrt{ {\tilde r}^2 + ( \kappa + p )^2 } + \sqrt{ {\tilde
r}^2 + ( \kappa - p )^2 } , \\
	&	&	\nonumber	\\
2 p {\bar y} & = & \sqrt{ {\tilde r}^2 + ( \kappa + p )^2 } - \sqrt{ {\tilde
r}^2 + ( \kappa - p )^2 } ,
\end{eqnarray}
\end{subequations}
where ${\tilde r} = r/m$ and $\kappa = b/m$.

In Fig. \ref{fig:disker} we show $\tilde D$ for  Kerr disks with different
values of $p$ and $\kappa$. We first plot $\tilde D$ for  disks with
$\kappa = 1.5$ and $p = 1$, $0.9$, $0.8$, $0.6$, and $0.3$. Then we show $\tilde
D$ for disks with $p = 0.8$ and $\kappa = 1.7$, $2$, $2.3$, $2.9$, and
$3.8$. We also  computed $\tilde D$ for different values of $p$ and $\kappa$,
we find that for some values of $p$ and $\kappa$, \; ${\tilde D} > 0$ everywhere on
the disk, whereas for the complementary  case $\tilde D$ have two roots, $0 <
{\tilde r}_1 < {\tilde r}_2$, so that ${\tilde D} < 0$ when 
$0 < {\tilde r}_1 <
{\tilde r}_2$. Accordingly we have two kinds of Kerr disks, ones
with heat flow in an  annular region ($0 < {\tilde r}_1 <
{\tilde r}_2$) and others without heat flow.

We first analyze some cases when there is no heat flow. In Fig.
\ref{fig:depre} we show the density ${\tilde \sigma} = m \sigma$ and the
pressure ${\tilde P} = m P$ for these Kerr disks. First we plot $\tilde \sigma$
and $\tilde P$ (scaled by a factor  $100$) for a  disk with $p = 0.9$ and
$\kappa = 2.4$, $2.5$, and $3$ as functions of ${\tilde r} = r/m$. Then we plot
$\tilde \sigma$ and $\tilde P$ (scaled by a factor  $10$) for  disks
with $\kappa = 2.5$ and $p = 1$, $0.9$, and $0.8$. We also computed $\tilde
\sigma$ and $\tilde P$ for a variety of  $p$ and $\kappa$, we found
a similar behaviour. The density $\tilde \sigma$ is positive
definite everywhere on the disk, falling to zero at infinity.
 There are values of $p$ and $\kappa$ such that the pressure $\tilde P$ is
positive definite, also falling to zero at infinity, whereas in other cases the
pressure $\tilde P$ is negative (tension) at the central region of the disks 
 becoming positive for a value ${\tilde r}_0 > 0$; then
falling to zero at infinity.

The rotational motion of these disks models is shown in Fig. \ref{fig:vemaker}.
We first plot $V$ for $p = 0.9$ with $\kappa = 2.4$, $2.5$, and $3$ and for $p =
0.8$ with $\kappa = 2.5$ (top curve) and then we plot the  disk angular
momentum $h$ for the same values of $p$ and $\kappa$. We can see that the disk
velocity  increases initially and then
falls to zero at infinity and always is less than the light velocity. Also 
we see that the disk angular momentum is an
increasing monotonic function of $\tilde r$. Thus these disks models are
stable.

Finally, we analyze a Kerr disk with heat flow, with $p = 0.9$ and $\kappa =
1.5$. In Fig. \ref{fig:dpcman} we show $\tilde \sigma$, $\tilde P$ and $\tilde
K$ for this values of $p$ and $\kappa$. We see that the density $\tilde \sigma$
is positive definite everywhere on the disk, falling to zero at infinity and
that presents a change of slope at the values of $\tilde r$ when ${\tilde D}
= 0$, ${\tilde r}_1$ and ${\tilde r}_2$. The pressure $\tilde P$ is negative at
the central region of the disk, becoming positive for a value of ${\tilde r} >
{\tilde r}_2$. Also, $\tilde P$ presents a change of slope  at ${\tilde r}_1$
and ${\tilde r}_2$. The heat flow function  $\tilde K$ increases rapidly and becomes
larger than $\tilde \sigma$ for a value of ${\tilde r} > {\tilde r}_1$, 
$\tilde K$ falls to zero, becoming less than $\tilde \sigma$ for a value of
${\tilde r} < {\tilde r}_2$. So, there is not causal propagation of heat on
this region of the disk.

We also plot in Fig. \ref{fig:dpcman} the disk angular momentum, $h$. We 
see that the angular momentum presents two instabilities, at the values of
$\tilde r$ when ${\tilde D} = 0$, ${\tilde r}_1$ and ${\tilde r}_2$, and so the
region of the disk where there is heat flow is highly instable. In order to
see that these instabilities are present also in the counter-rotating model,
we plot also the stream angular momenta, $h_+$ and $h_-$ (scaled by a factor
 $20$). We analyzed several other cases of Kerr disks with heat flow, we found 
the same behaviour.

\section{Discussion}

We presented two families of rotating disks based on the Taub-NUT and Kerr
solutions. For the first family we find that the presence of heat flow is a
generic property of  this models, also  generic is the radial instability 
produced by this flow. For the second family we find  two sub classes one
with and  the other without heat flow. The  heat flow is concentrated 
in an  annular region that is highly instable. 

The relation found between the geodesic circular velocities and the ``speed of sound"
found  is a generalization for the stationary case of the corresponding relation
found by Morgan and Morgan \cite{MM2} that is the base of the counter-rotating
model. 

The inclusion of current in the rotating models to have a rotating ``hot" disk 
as the ones studied in \cite{hot} is under consideration. The rotation add mass
to the otherwise massless charges. These models are quite nontrivial. Also, the
inclusion of radial pressure or tension is being considered. 

Finally, we want to mention that all the computation of this work was 
 performed
using the algebraic programming system Reduce \cite{reduce}.

\vspace{0.3cm}
\noindent
{\bf Acknowledgments}

We want to thank  CNPq   and FAPESP for financial support. Also
 G.A.G  is grateful for the warm hospitality of 
 the DMA-IMECC-UNICAMP where the main part of this work was performed.


\newpage

\begin{figure}
$$\begin{array}{cc}
{\tilde D} \ ; \ \kappa = 1.7 & {\tilde D} \ ; \ p = 0.8 \\
\epsfig{width=2.5in,file=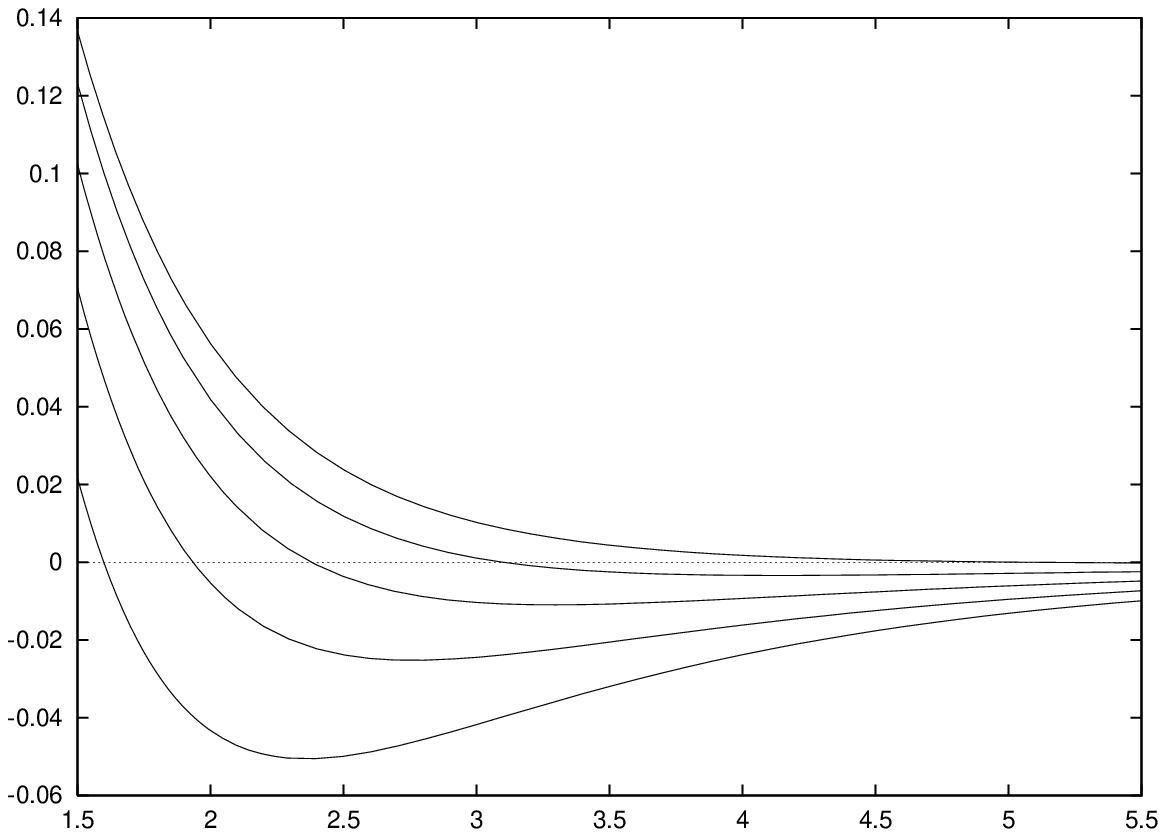} &
\epsfig{width=2.5in,file=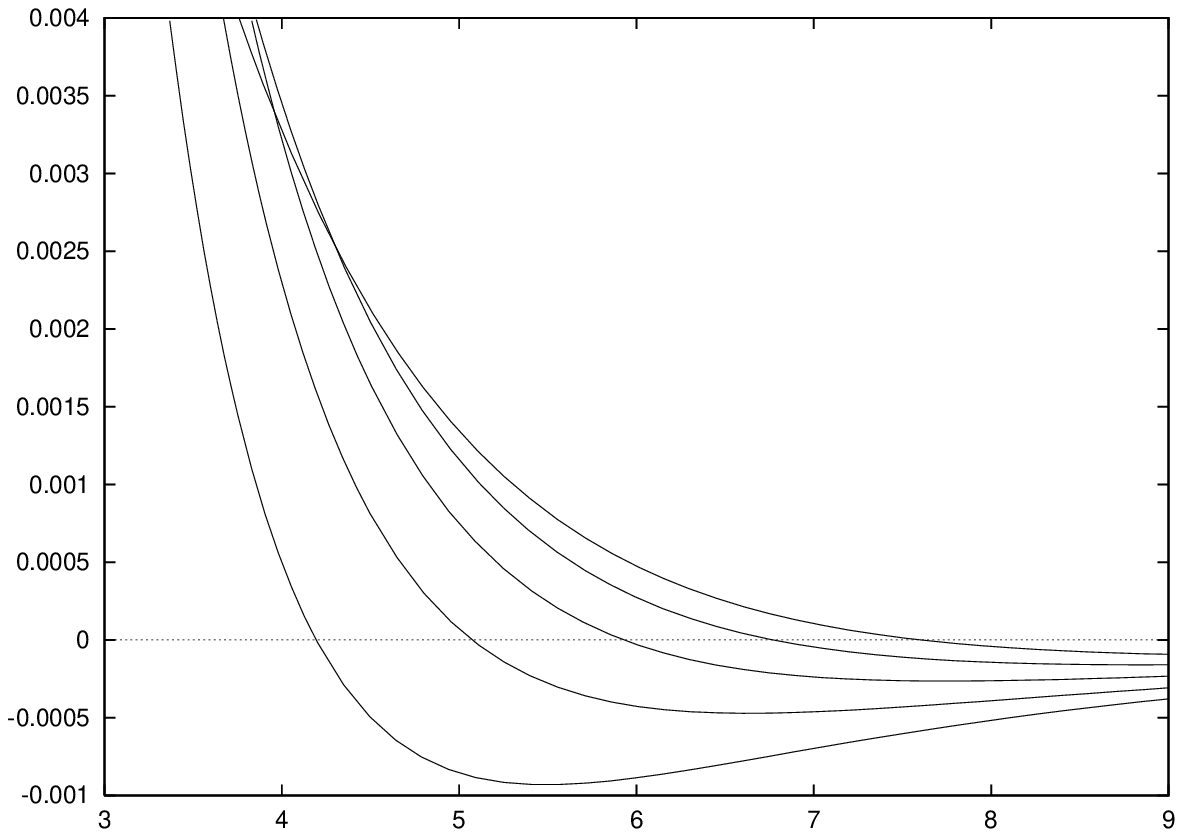} \\
{\tilde r} = r/\alpha & {\tilde r} = r/\alpha
\end{array}$$
\caption{We plot $\tilde D$ for Taub-NUT disks with $\kappa = 1.7$ and $p
= 0.9$, $0.7$, $0.5$, $0.3$, and $0.1$ as functions of ${\tilde r} = r/\alpha$.
Next we plot $\tilde D$ for Taub-NUT disks with $p = 0.8$ and $\kappa = 2$,
$2.6$, $3.2$, $3.8$, and $4.4$ as functions of ${\tilde r} =
r/\alpha$. Note that $D$ is not positive definite and
has only one root ${\tilde r}_0 > 0$   }\label{fig:disnut}
\end{figure}

\begin{figure}
$$\begin{array}{cc}
{\tilde \sigma} \ , \ {\tilde P} \ ; \ \kappa \ = \ 1.4 &
{\tilde \sigma} \ , \ {\tilde P} \ ; \ \kappa \ = \ 1.7 \\
\epsfig{width=2.5in,file=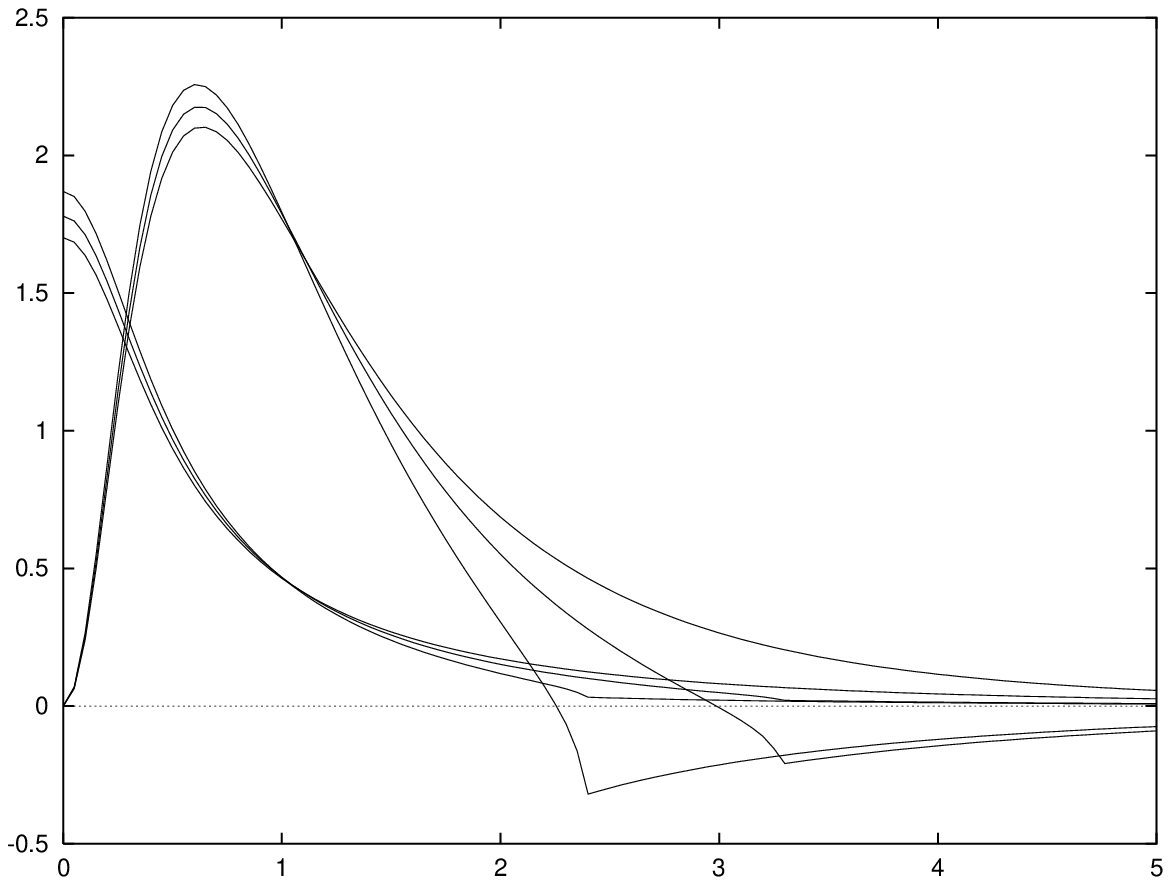} &
\epsfig{width=2.5in,file=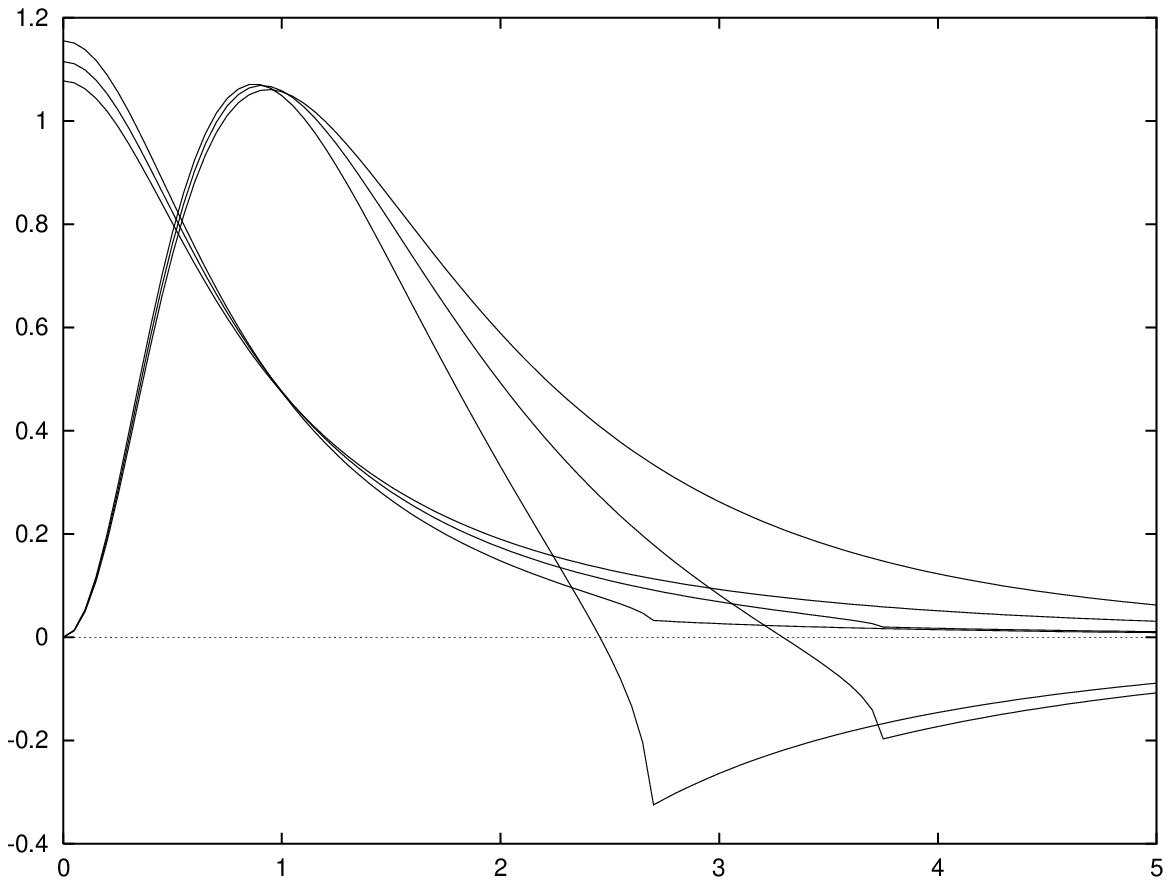} \\
{\tilde r} = r/\alpha & {\tilde r} = r/\alpha
\end{array}$$
\caption{First we plot $\tilde \sigma$ and $\tilde P$ (scaled by a factor
$10$) for Taub-NUT disks with $\kappa = 1.4$ and $p = 1$, $0.8$, and $0.6$ as
functions of $\tilde r = r/m$. Next we plot $\tilde \sigma$ and $\tilde P$
(scaled by a factor $10$) for a Taub-NUT disk with $\kappa = 1.7$ and $p = 1$,
$0.8$, and $0.6$ as functions of $\tilde r = r/m$. $\tilde P$ presents 
a non smooth behaviour at
${\tilde r} = {\tilde r}_0$, wherein the heat flow begins.
}\label{fig:depenut}
\end{figure}

\begin{figure}
$$\begin{array}{cc}
{\tilde K} & {\tilde \sigma} \ , \ {\tilde P} \ , {\tilde K} \\
\epsfig{width=2.5in,file=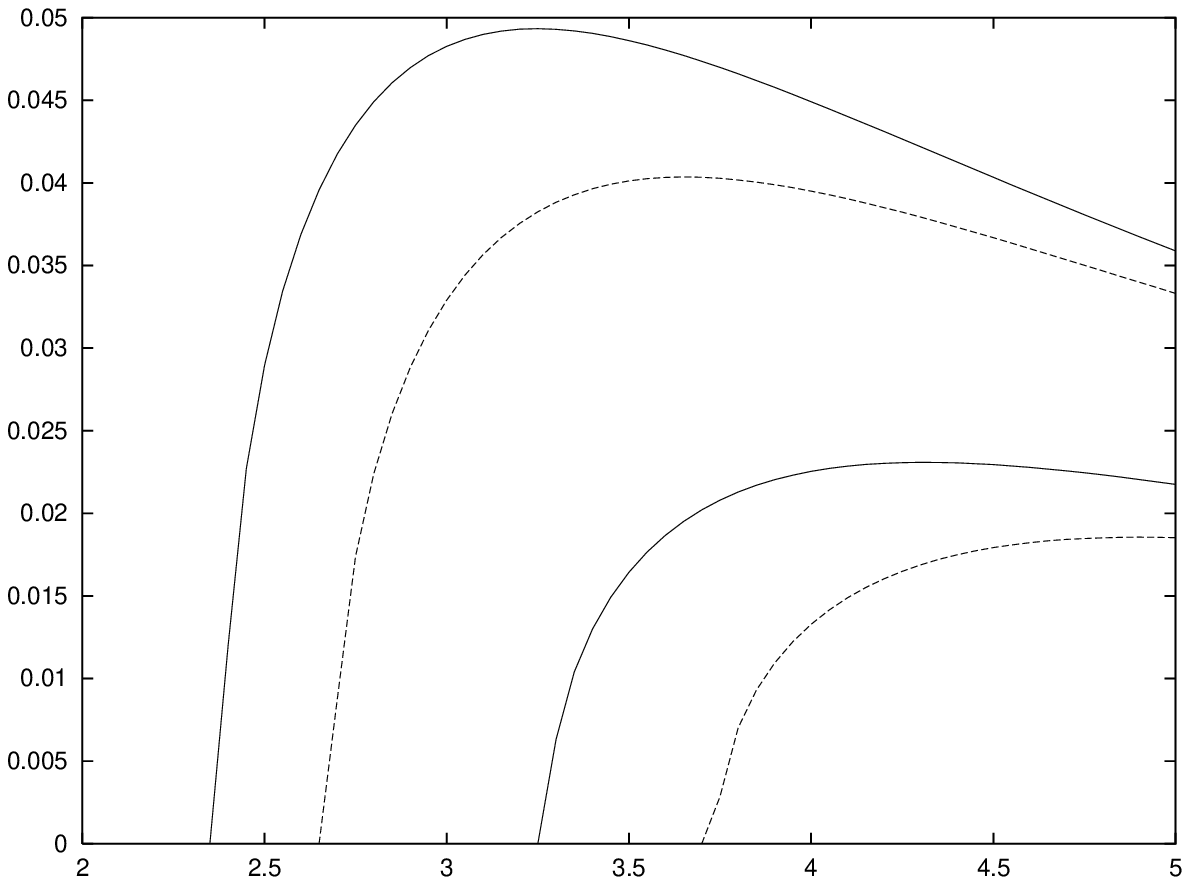} &
\epsfig{width=2.5in,file=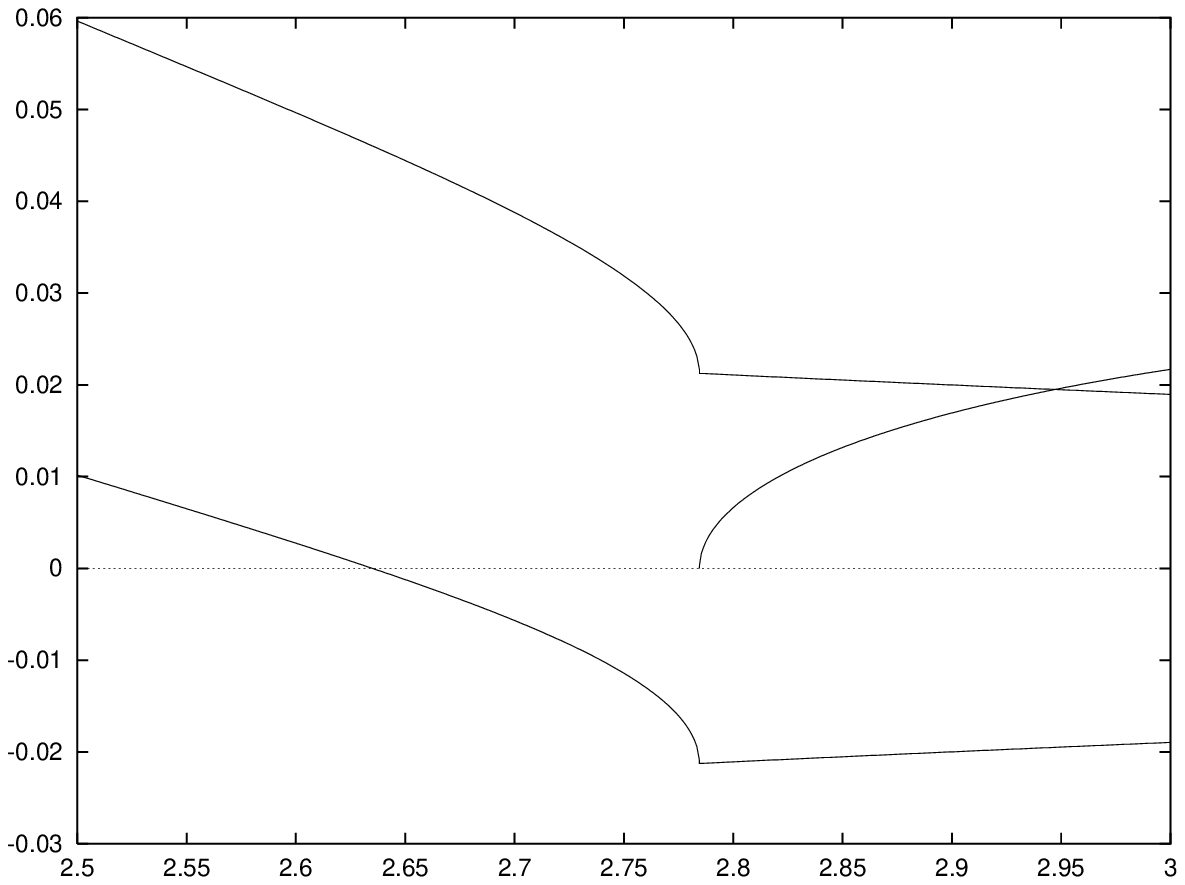} \\
{\tilde r} = r/\alpha & {\tilde r} = r/\alpha 
\end{array}$$
\caption{We show  $\tilde K$ for $\kappa = 1.4$ (upper curves) and $\kappa =
1.7$ (lower curves), with $p = 0.8$ and $0.6$ as functions of
$\tilde r$. 
Next we plot $\tilde \sigma$, $\tilde P$ and $\tilde K$ for 
$\kappa = 1.1$ and $p =
0.8$ in the interval $2.5 \leq {\tilde r} \leq 3$.}\label{fig:calnut}
\end{figure}

\begin{figure}
$$\begin{array}{cc}
V & h \ , \ h_+ \ , \ h_-  \\
\epsfig{width=2.5in,file=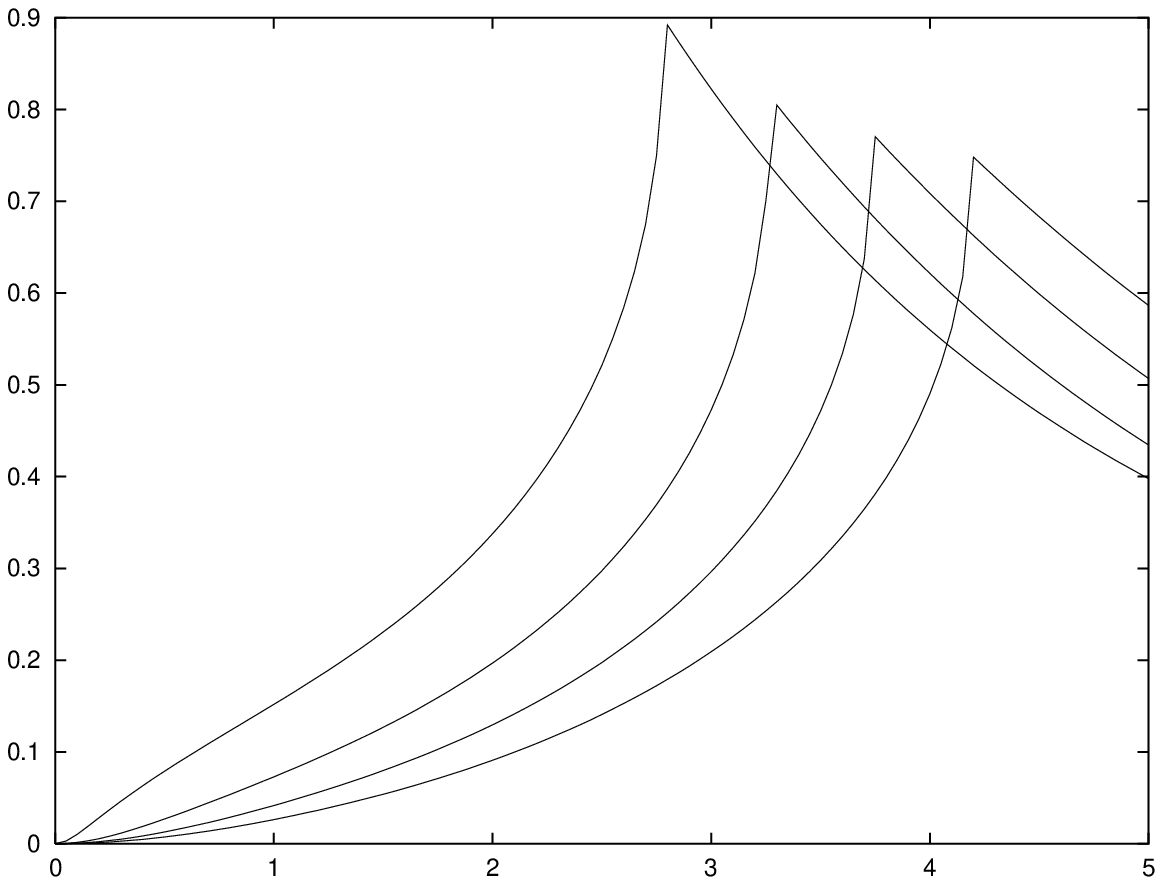} &
\epsfig{width=2.5in,file=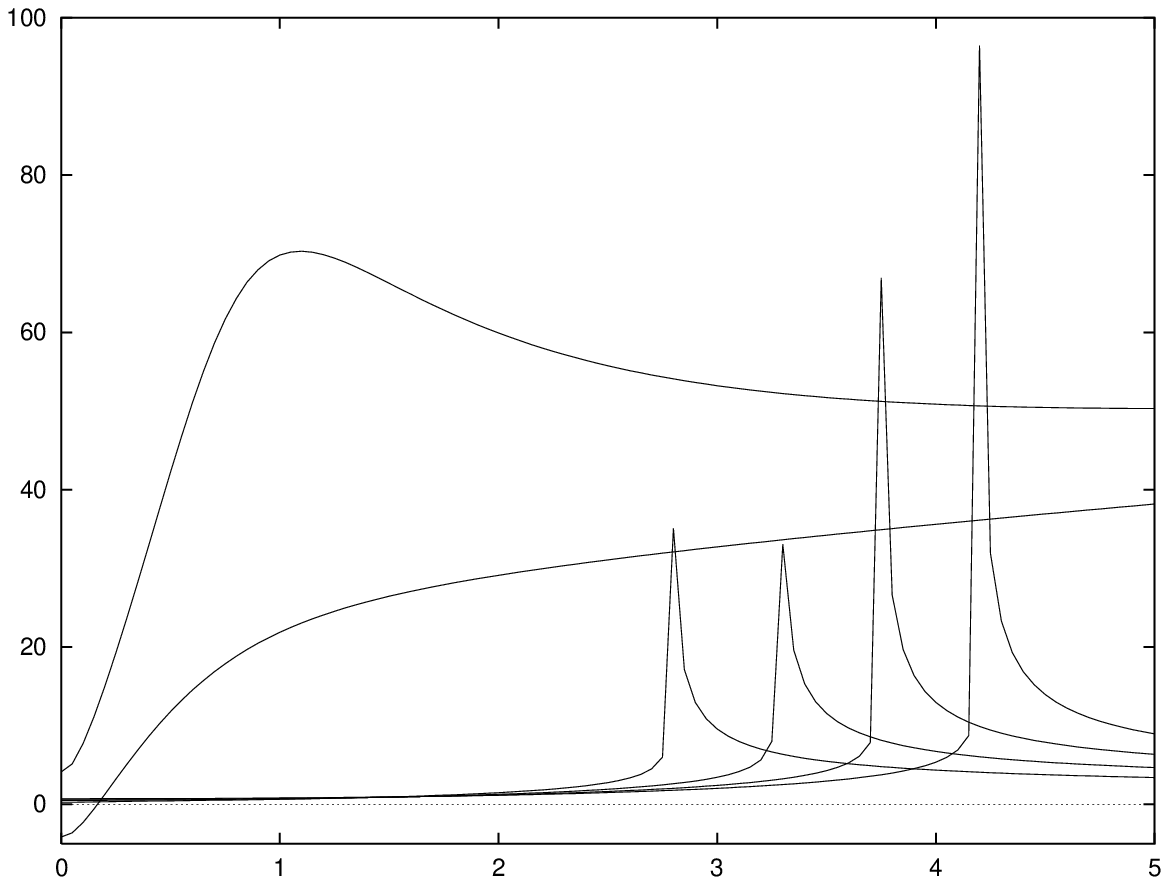} \\
{\tilde r} = r/\alpha & {\tilde r} = r/\alpha
\end{array}$$
\caption{We plot the  tangential disk velocity,  $V$, with $p = 0.8$ and 
$\kappa = 1.1$, $1.4$, $1.7$, and $2$ as functions of $\tilde r$. Next we plot, 
 for the same values of $p$ and $\kappa$ ,
the disk angular momentum, $- h$ (sharp curves), and
the stream angular momenta, $h_+$ (top curve from the left) and $- h_-$ 
(scaled by a factor of
$15$), for $p = 0.8$ and $\kappa = 1.1$ as functions of $\tilde
r$.}\label{fig:vemanut}
\end{figure}

\begin{figure}
$$\begin{array}{cc}
U_+, \ U_-, \ U_+U_-, \ P/\sigma & P/\sigma \\
\epsfig{width=2.5in,file=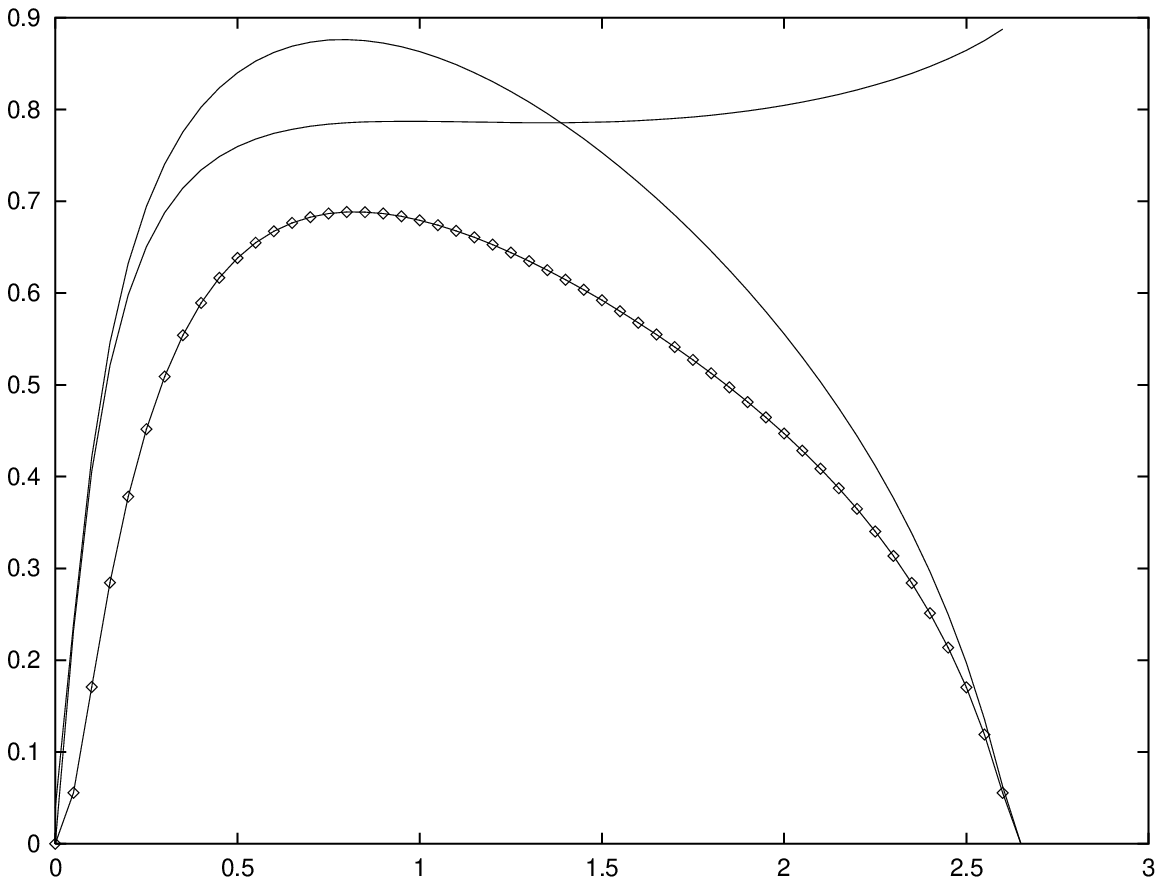} &
\epsfig{width=2.5in,file=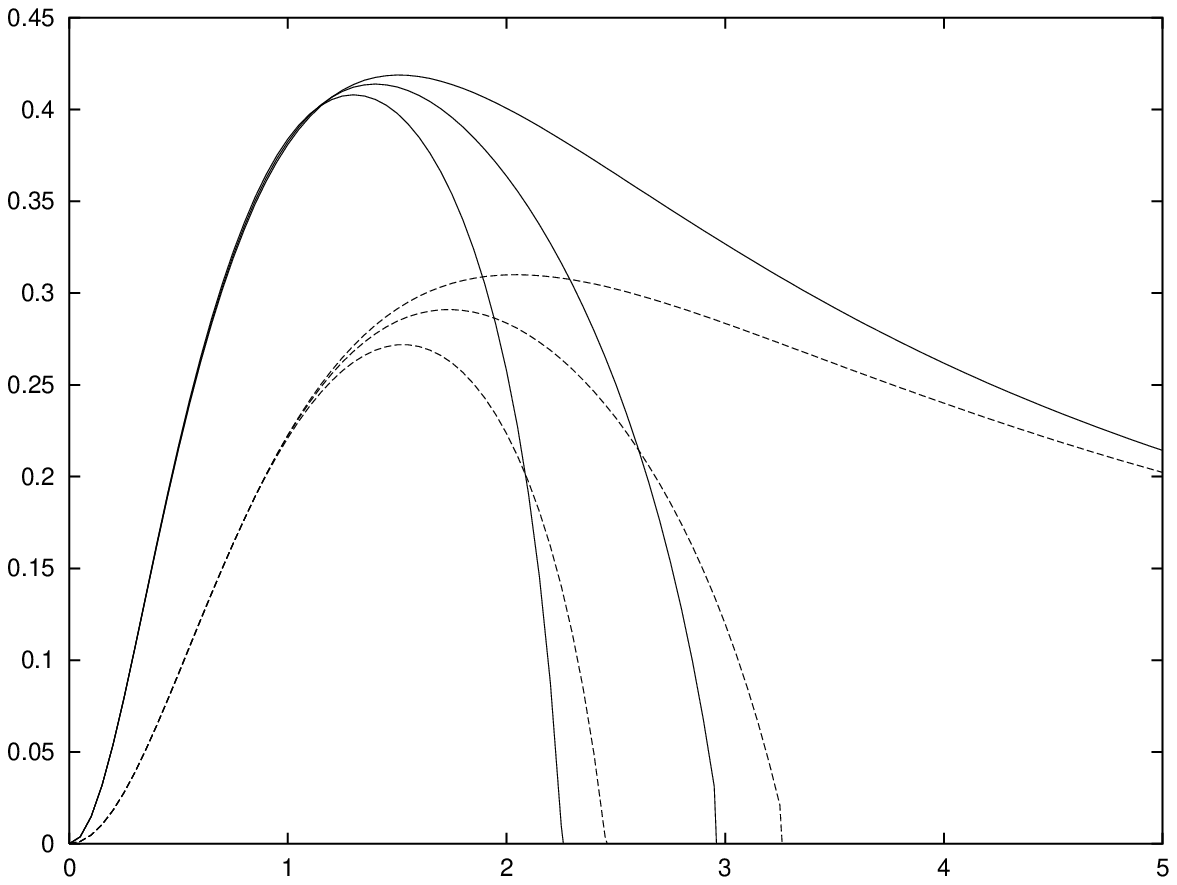} \\
{\tilde r} = r/\alpha & {\tilde r} = r/\alpha
\end{array}$$
\caption{We show $U_+$, $U_-$, $U_+U_-$ and $P/\sigma$ (full dots) for
$\kappa = 1.1$ and $p = 0.8$ as functions of $\tilde r$. Next we plot the
$P/\sigma$ relation for $\kappa = 1.4$ (upper curves) and $\kappa = 1.7$ (lower
curves), with $p = 1$, $0.8$, and $0.6$ as functions of $\tilde
r$.}\label{fig:vcr}
\end{figure}

\begin{figure}
$$\begin{array}{cc}
{\tilde D} \ ; \ \kappa = 1.5 & {\tilde D} \ ; \ p = 0.8 \\
\epsfig{width=2.5in,file=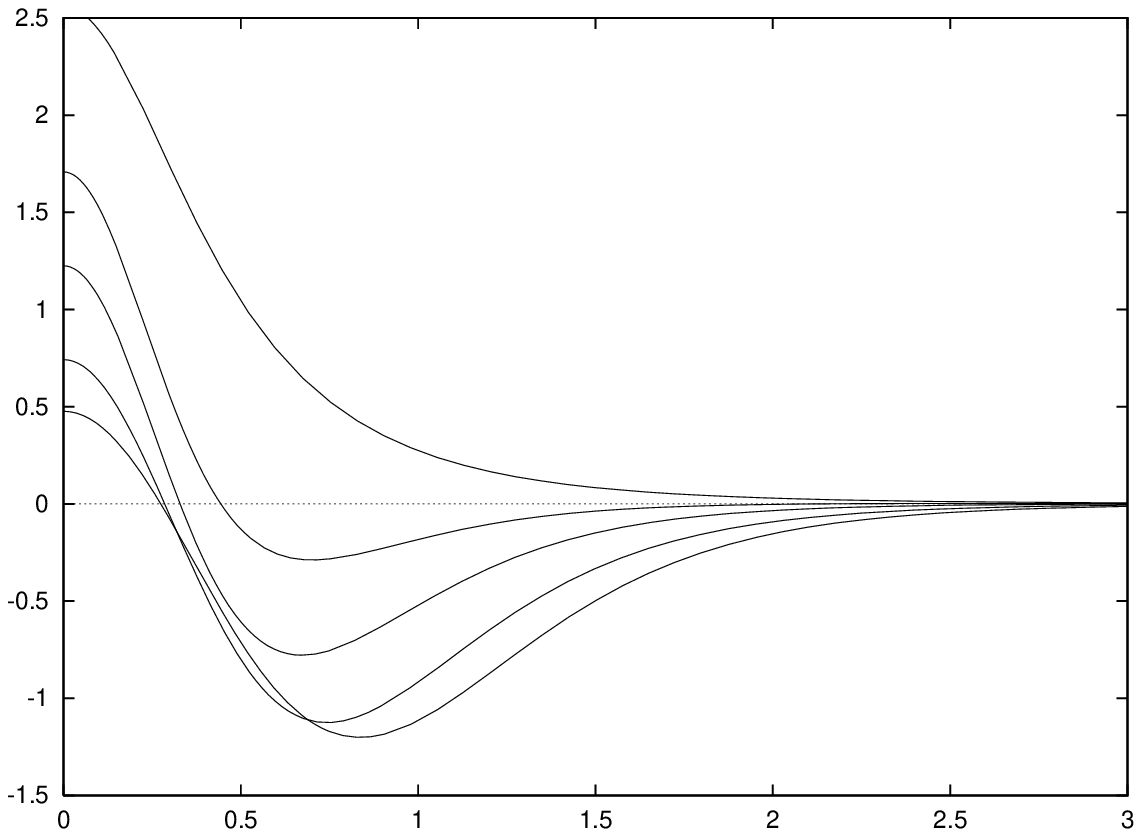} &
\epsfig{width=2.5in,file=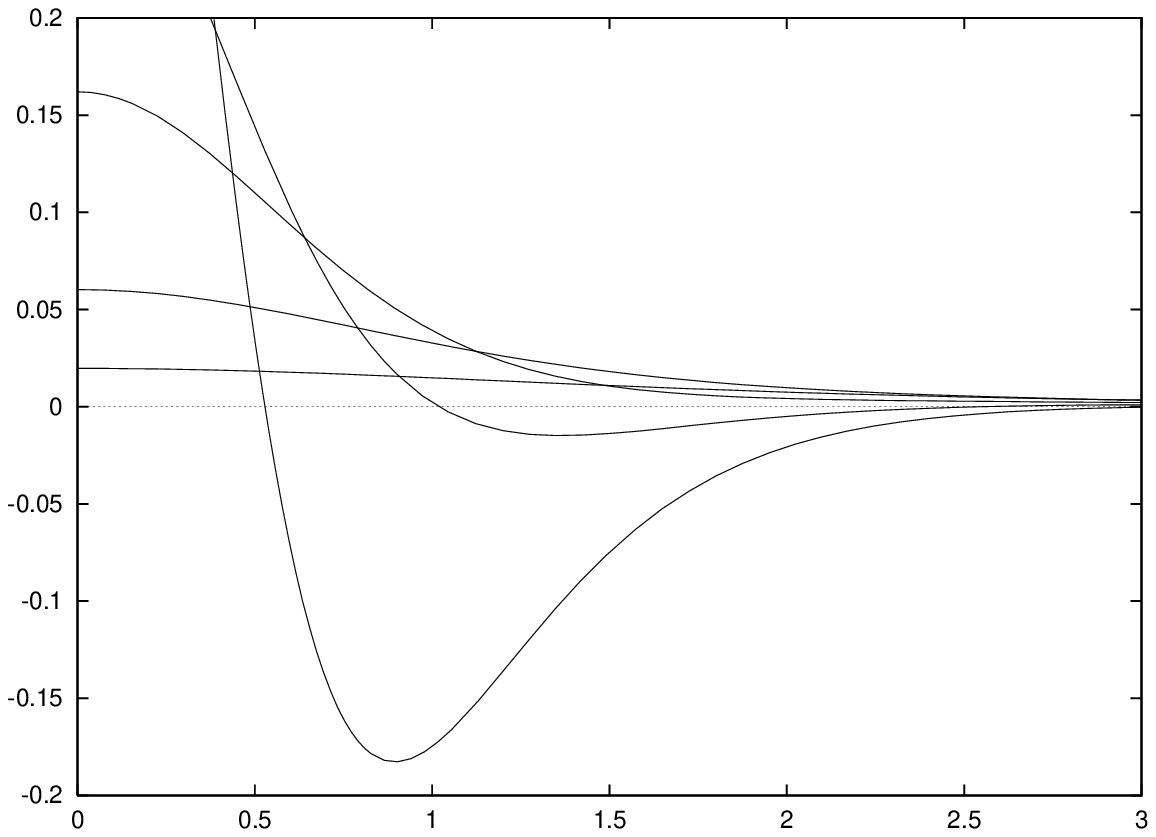} \\
{\tilde r} = r/m & {\tilde r} = r/m 
\end{array}$$
\caption{The discriminant $\tilde D$ associated with the 
  Kerr disks with $\kappa = 1.5$ is
plotted for  $p = 1$, $0.9$, $0.8$, $0.6$, and $0.3$.  Next we show
$\tilde D$ for  disks with $p = 0.8$  and  $\kappa = 1.7$, $2$,
$2.3$, $2.9$ and $3.8$. Note that $\tilde D$ is not positive defined. }\label{fig:disker}
\end{figure}

\begin{figure}
$$\begin{array}{cc}
{\tilde \sigma} \ , \ {\tilde P} \ ; \ p \ = \ 0.9 &
{\tilde \sigma} \ , \ {\tilde P} \ ; \ \kappa \ = \ 2.5 \\
\epsfig{width=2.5in,file=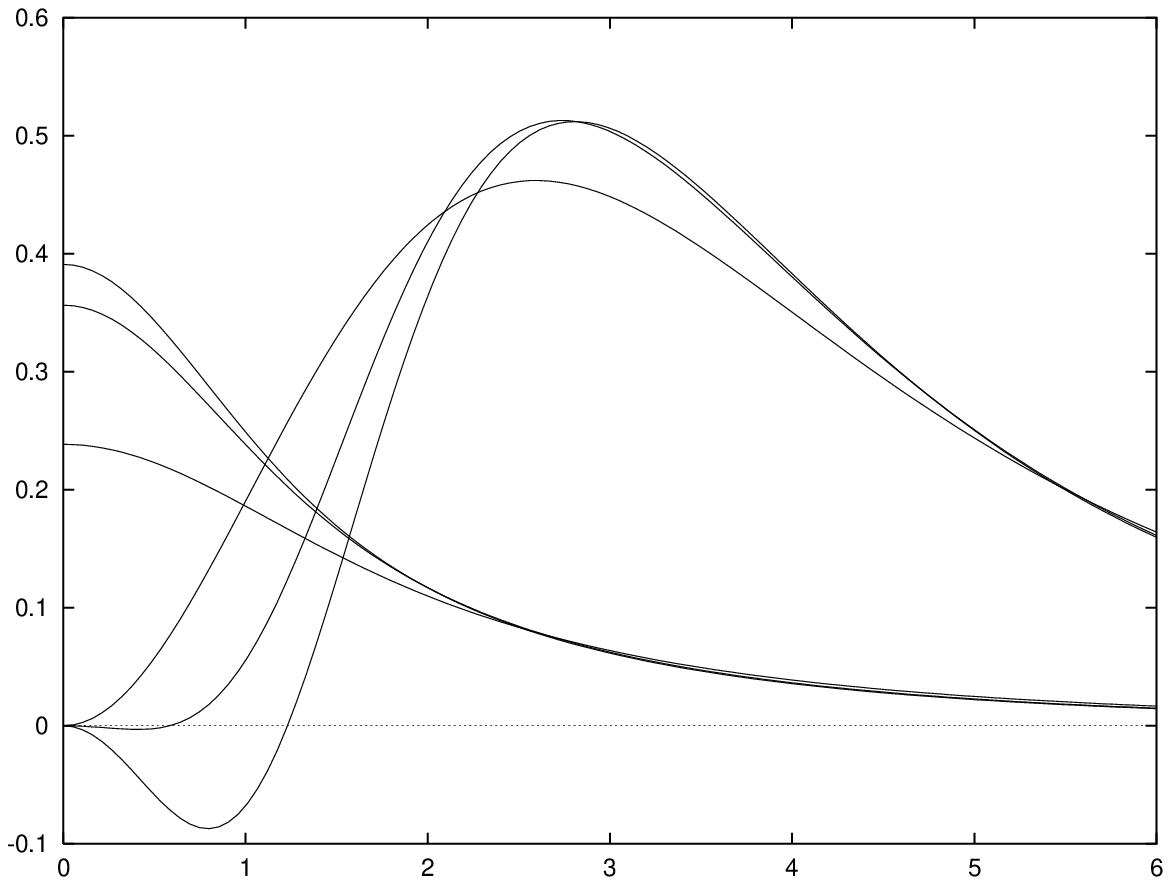} &
\epsfig{width=2.5in,file=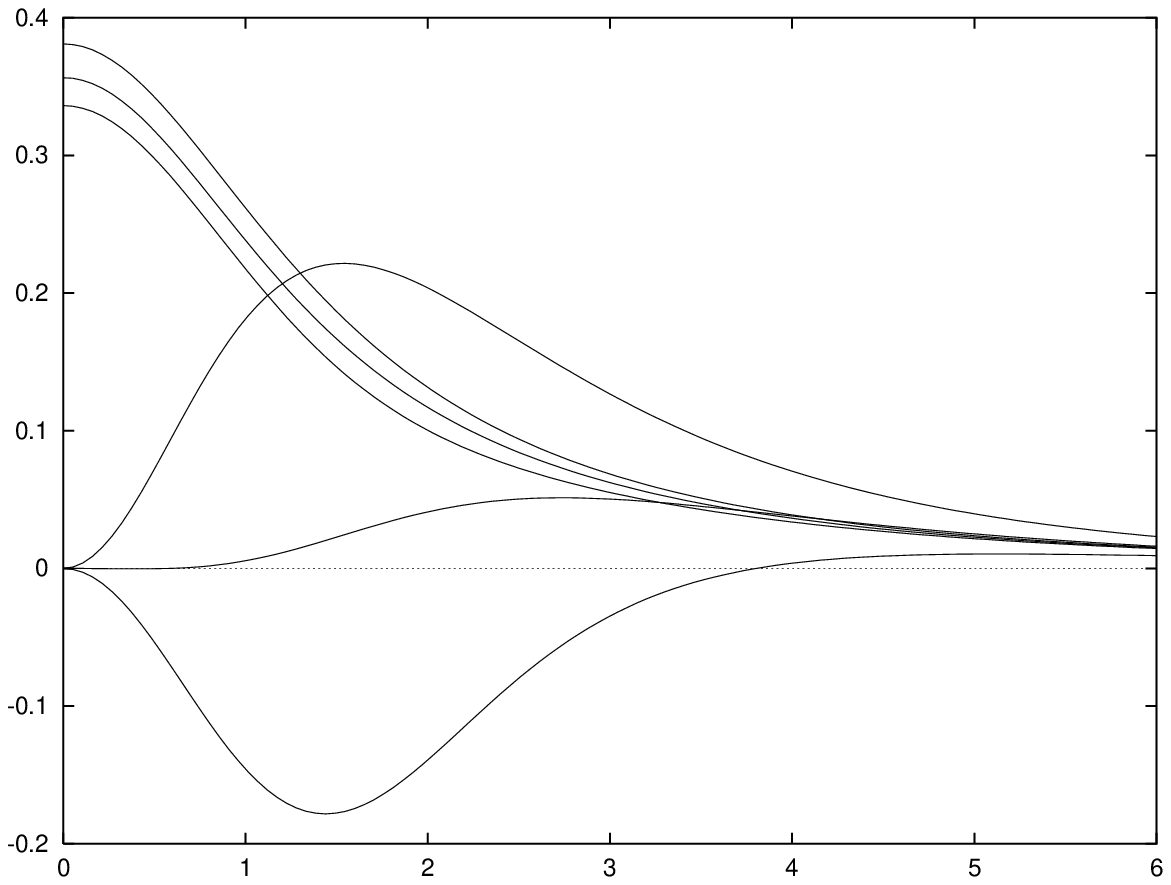} \\
{\tilde r} = r/m & {\tilde r} = r/m \\
\end{array}$$
\caption{First we plot $\tilde \sigma$ and $\tilde P$ (scaled by a factor of
$100$) for  Kerr disks with $p = 0.9$ and $\kappa = 2.4$, $2.5$, and $3$ as
functions of ${\tilde r} = r/m$. Next we plot $\tilde \sigma$ and $\tilde P$
(scaled by a factor of $10$) for  disks with $\kappa = 2.5$ and $p = 1$,
$0.9$ and $0.8$.}\label{fig:depre}
\end{figure}

\begin{figure}
$$\begin{array}{cc}
V \ ; \ p = 0.9 & h \ ; \ p = 0.9 \\
\epsfig{width=2.5in,file=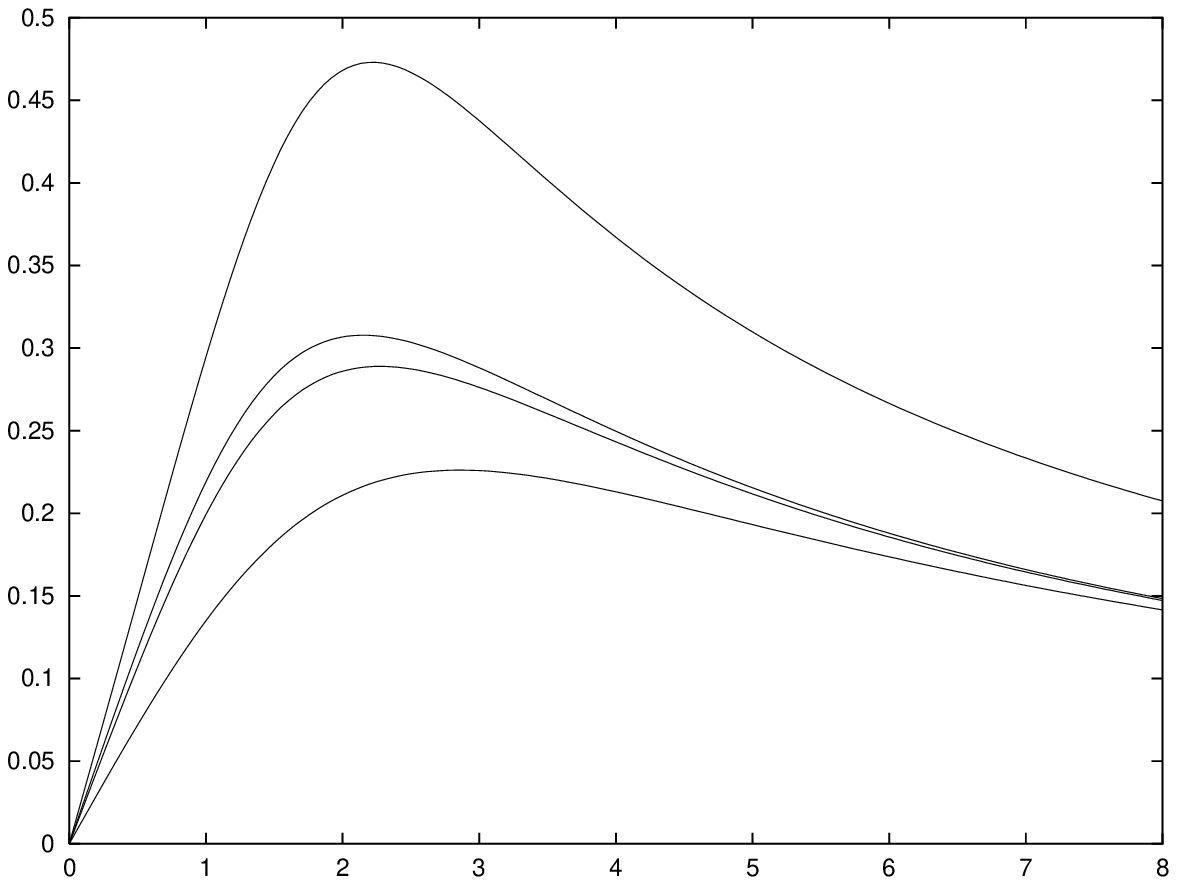} &
\epsfig{width=2.5in,file=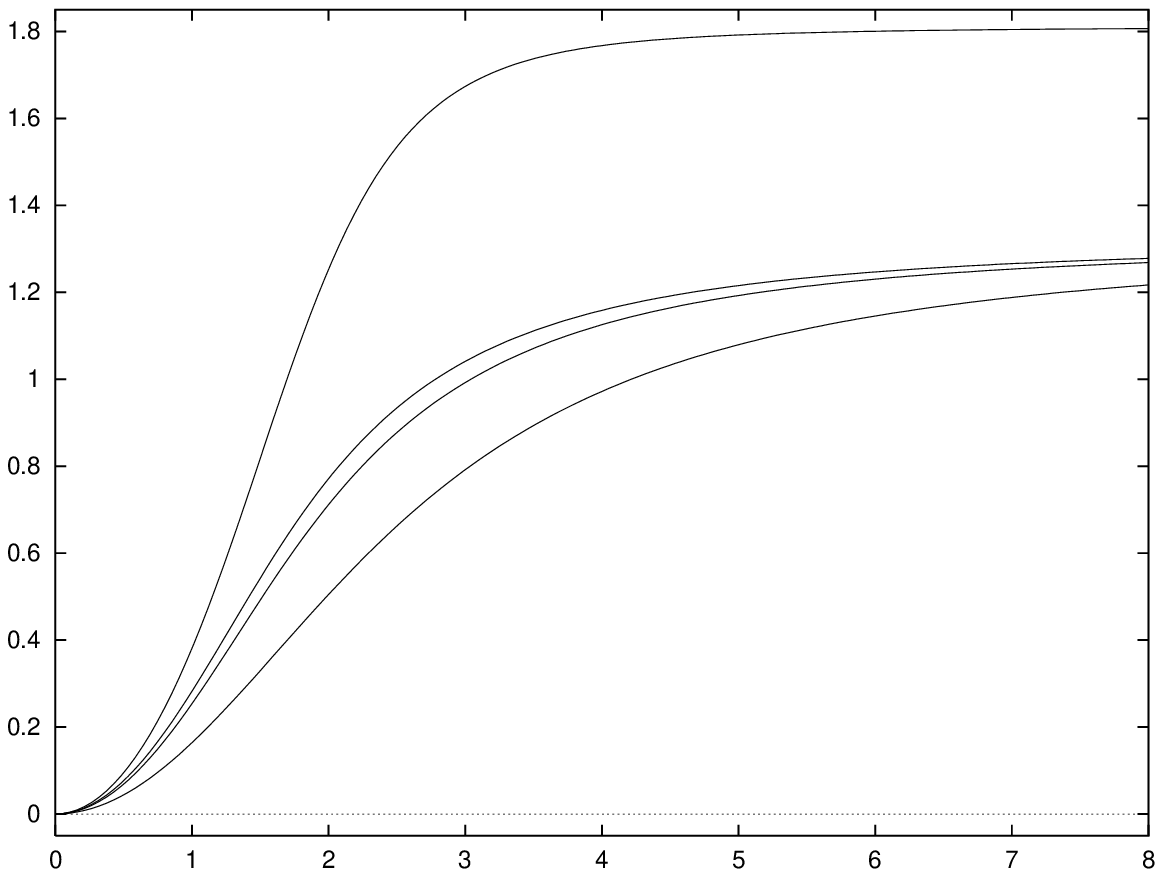} \\
{\tilde r} = r/m & {\tilde r} = r/m
\end{array}$$
\caption{We plot $V$ for $p = 0.8$ with $\kappa = 2.5$ (top curve) and  
$p = 0.9$ with $\kappa = 2.4$, $2.5$,and $3$ (following three curves). 
Next we plot the disk
angular momentum, $h$. for the same values of $p$ and
$\kappa$.}\label{fig:vemaker}
\end{figure}

\begin{figure}
$$\begin{array}{cc}
\sigma \ , P \ , K & h \ , \ h_+ \ , \ h_- \\
\epsfig{width=2.5in,file=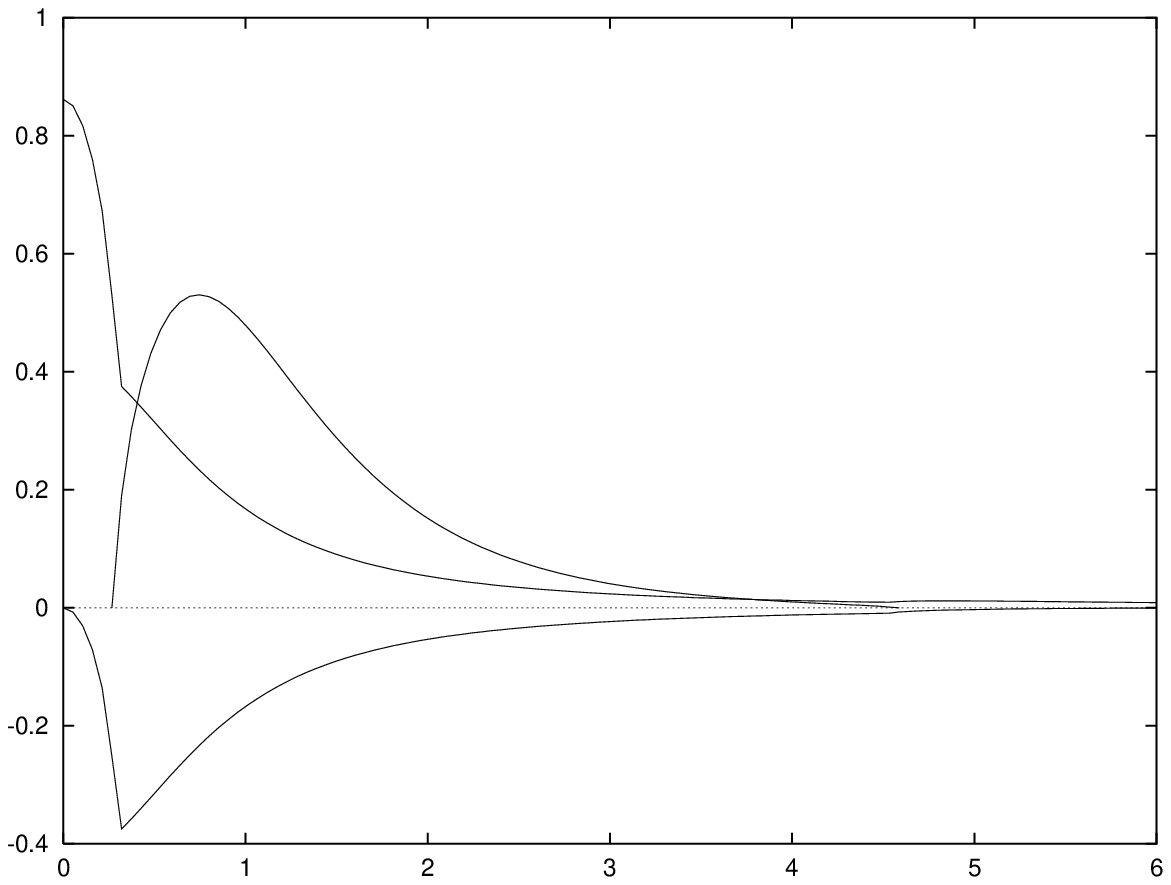} &
\epsfig{width=2.5in,file=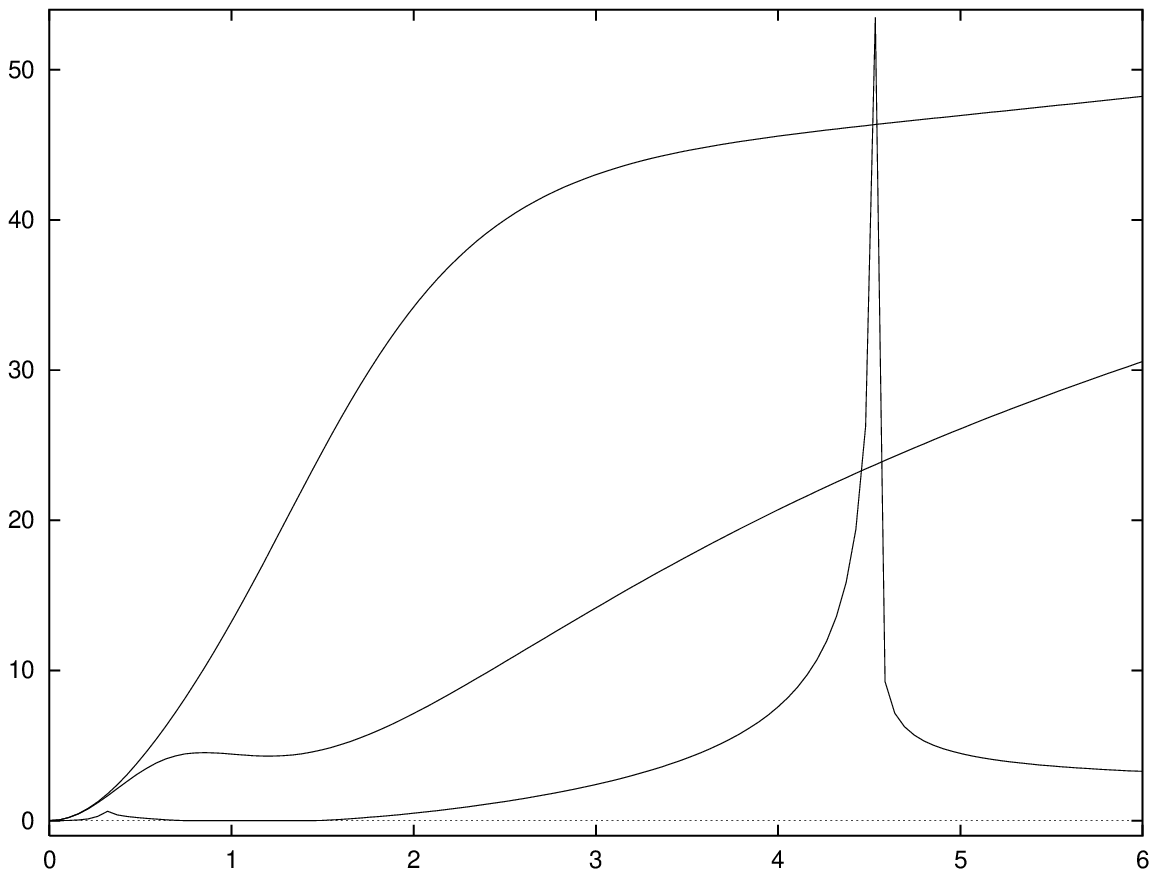} \\
{\tilde r} = r/m & {\tilde r} = r/m
\end{array}$$
\caption{We plot the density , $\tilde \sigma$, the pressure, $\tilde P$, 
and the heat function function, $K$, for  a Kerr disk with $p = 0.9$ and $\kappa = 1.5$ as functions
of ${\tilde r} = r/m$. Next, for the same values of $p$ and $\kappa$, we plot
 the disk
angular momentum, $h$, and the stream angular momenta, $h_+$ and $h_-$ (scaled
by a factor  $20$).}\label{fig:dpcman}
\end{figure}

\end{document}